\documentclass[namedreferences]{solarphysics}
\usepackage[optionalrh]{spr-sola-addons} 
\usepackage{graphicx}        
\usepackage{natbib}         
\usepackage{color}           
\usepackage{url}             

\newcommand{\rsun}{$R_\odot$}

\newcommand{\adv}{    {\it Adv. Space Res.}}

\newcommand{\aap}{    {\it Astron. Astrophys.}}

\newcommand{\apj}{    {\it Astrophys. J.}}

\newcommand{\grl}{    {\it Geophys. Res. Lett.}}

\newcommand{\jgr}{    {\it J. Geophys. Res.}}

\newcommand{\solphys}{{\it Solar Phys.}}

\newcommand{\ssr}{    {\it Space Sci. Rev.}}

\begin{document}

\begin{article}

\begin{opening}

\title{Coronal Pseudo-Streamer and Bipolar Streamer Observed by SOHO/UVCS in March 2008}

\author{L.~\surname{Abbo}$^{1}$\sep
        R.~\surname{Lionello}$^{2}$\sep
        P.~\surname{Riley}$^{2}$\sep
        Y.-M.~\surname{Wang}$^{3}$
   }   
\runningauthor{Abbo et al.}
\runningtitle{Coronal Pseudo and Bipolar Streamers}

   \institute{$^{1}$ INAF-Osservatorio Astrofisico di Torino, Pino Torinese 10025, Italy
                     email: abbo@oato.inaf.it\\
              $^{2}$ Predictive Science, Inc., 9990 Mesa Rim Road, San Diego, CA 92121, USA\\
              $^{3}$ Space Science Division, Naval Research Laboratory, Washington, DC 20375-5352, USA\\
             }

\begin{abstract}
The last solar minimum is characterized by several peculiar aspects and by the presence of a complex magnetic topology with two different kinds of coronal streamers: pseudo-streamers and bipolar streamers.
Pseudo-streamers or unipolar streamer are coronal structures which separate coronal holes of the same polarity, without a current sheet in the outer corona; unlike bipolar streamer that separate coronal holes of opposite magnetic polarity. In this study, two examples of these structures have been identified in the period of Carrington rotation 2067, by applying a potential-field source-surface extrapolation of the photospheric field measurements.
We present a spectroscopic analysis of a pseudo-streamer and a bipolar streamer observed in the period 12-17 March 2008  at high spectral and
spatial resolution by the {\it Ultraviolet Coronagraph Spectrometer} (UVCS; Kohl {\it et al.}, 1995)
 onboard {\it Solar and Heliospheric Observatory} (SOHO).
The solar wind plasma parameters, such as kinetic temperature, electron density
and outflow velocity, are inferred
in the extended corona (from 1.7 to 2.1 \rsun) analysing the O {\sc vi} doublet and  Ly $\alpha$ line spectra. 
 The coronal magnetic topology is taken into account and has been
extrapolated by a 3D magneto-hydrodynamic model of the global corona.
The results of the analysis show some peculiarities of the pseudo-streamer physical parameters in comparison with those obtained for bipolar streamers: in particular, we have found higher kinetic temperature and higher outflow velocities of O {\sc vi} ions and lower electron density values. In conclusion, we point out that pseudo-streamers produce a ''hybrid'' type of outflow that is intermediate between slow and fast solar wind and they are a possible source of slow/fast wind in not dipolar solar magnetic field configuration.
\end{abstract}
\end{opening}

\section{Introduction}
The last solar minimum at the end of solar cycle 23 has been characterized by several aspects, very peculiar and different in comparison with the previous minimum. Polar coronal holes appear smaller, and polar magnetic flux measured at the solar surface is 40$\%$ weaker relative to 1996 minimum (Kirk {\it et al.}, 2009), magnetic fields measured in the solar wind above the Sun's poles are depleted by a similar amount (about a third;  Smith and Balogh, 2008), solar wind density is decreased (by 17-20$\%$), and the bulk solar wind speed is slightly slower ($\approx 3\%$; McComas {\it et al.}, 2008). This period has been exceptionally quiet, with sunspot numbers at their lowest in a century and the presence of low-latitude large coronal holes implies strong, long and recurring high speed streams in the solar wind at the Earth's orbit (Gibson {\it et al.}, 2009).
 Moreover, the coronal magnetic topology is rather unexpected for a solar minimum based on what we had observed in other space-age minima. Instead of the classical dipolar configuration with a single equatorial belt, the magnetic structure is more complex and, in particular, it is characterized by the presence of two different kinds of coronal streamers called pseudo-streamers and bipolar streamers (Riley {\it et al.}, 2011).
Pseudo-streamers or unipolar streamer are coronal structures which separate coronal holes of the same polarity, overlying twin loop arcades without a current sheet in the outer corona ({\it e.g.} Zhao and Webb 2003; Wang {\it et al.}, 2007). 
Bipolar streamers or helmet streamer overlie a single (or an odd number of) loop arcades in the lower corona and they have oppositely oriented open magnetic field in the upper corona, such that a current sheet is present between the two open field domains.
Slow wind {\it in-situ} measurements associated with a pseudo-streamer indicate wind speed in the range of about 350 - 550 km s$^{-1}$ with oxygen charge-state ratio ($n_{{\rm O}^{7+}}$/n$_{{\rm O}^{6+}}$) greater than fast wind values and lower than slow wind values (Wang {\it et al.}, 2012). Zhao {\it et al.} (2013) analysed {\it Ulysses} observations of solar wind proton flux extremes originating from sources middle-distant from the heliospheric current sheet, likely related to pseudo-streamer structures. Moreover, interplanetary signatures derived by the {\it Advanced Composition Explorer} (ACE) data at 1 AU show that pseudo-streamer flows have all the same characteristics as slow wind flows, that is low speed and proton temperature and high density and composition ratio, but less pronounced in pseudo-streamers (Riley {\it et al.}, 2012; Crooker {\it et al.}, 2014). 
The analysis presented here concerns a coronal pseudo-streamer and a bipolar streamer observed in the period 12-17 March 2008 by SOHO/UVCS at high spectral and
spatial resolution, in order to compare the physical parameters of these coronal structures, such as kinetic temperature of H {\sc i} Ly $\alpha$ and O {\sc vi} spectral lines, electron density and outflow velocity. The diagnostic techniques that we apply to derive the electron density as a function of outflow velocity is described in Section 2, the UVCS observations and the identification of the coronal structures are described in Section 3 and the 3D magneto-hydrodynamic model used for the extrapolations of the coronal magnetic field is illustrated in Section 4. Finally, Sections 5 and 6 are dedicated to the data analysis and the discussion of the results, respectively.  

\section{Diagnostic Techniques for the Coronal Plasma}

We derive the outflow velocity and electron density of the coronal wind plasma from the
emission of the doublet O~{\sc vi} 1032 and 1038 {\AA} spectral lines observed by UVCS.
 These lines are formed
in the extended corona via collisional
 and radiative excitation
processes ({\it e.g.} Withbroe {\it et al.}, 1982). The two components of the O~{\sc vi} 1032 and 1038 {\AA} lines in an expanding plasma can be separated by
using the method introduced by Antonucci {\it et al.} (2004) and already applied in the analysis by Abbo {\it et al.} (2010). Because the two components have a
different dependence on the electron density (the collisional process depends
on $n_{\rm e}^2$, while the radiative process
  depends linearly on electron density $n_{\rm e}$), it results that the electron
density, averaged along the line-of-sight (LOS), $\langle< n_{\rm e}\rangle>$, is proportional to the ratio of the collisional component $I_{\rm c}$
 to the radiative component $I_{\rm r}$, and is a function of the outflow
velocity of the wind, {\bf w}, through the relationship $\langle n_{\rm e}\rangle\,\approx\,\frac{I_{\rm c}}{I_{\rm r}}\,\langle \Phi(\delta\lambda)\rangle$,
where $\langle \Phi(\delta\lambda)\rangle$  is the Doppler dimming function which depends
on the normalized
 coronal absorption profile and on the intensity of the exciting spectrum along the direction of the incident radiation,
$\bf n$. The quantity $\delta\lambda=\frac{\lambda_0}{c}\,{\bf w}\cdot{\bf n}$ is
 the shift of the disk spectrum
introduced by the expansion velocity, {\bf w}, of the coronal absorbing
ions/atoms along the direction $\bf n$, and $\lambda_0$ is the reference wavelength of the
transition. As the wavelength shift increases, the resonantly scattered emission decreases,
 giving origin to the Doppler dimming effect (Beckers and Chipman, 1974; Noci {\it et al.}, 1987). By analysing the
 O~{\sc vi} doublet lines at 1031.93 and
1037.62 \AA, it is
possible to measure oxygen ion outflow velocities (averaged along the LOS on the plane of sky) up to approximately 450 km s$^{-1}$ for
 the effect of pumping
of the C{\sc ii} lines at 1037.02 and 1036.34 \AA~on the O~{\sc vi} $\lambda$~1037.61 {\AA} line ({\it e.g.} Dodero {\it et al.}, 1998; Cranmer {\it et al.}, 1999).
 The disk intensity values of the O {\sc vi} and C {\sc ii} lines are obtained by Curdt {\it et al.} (2001).
When the plasma is dynamic, we need a further physical constraint which is  the mass flux conservation along the
flow tube connecting the corona to the heliosphere by taking into account the expansion factors of the flux tubes as derived by the MHD model (see the next section) and by considering the mass flux measured in the heliosphere
by {\it Ulysses} (McComas {\it et al.}, 2000, 2008).

\begin{table}[!h]
\caption{ Details for the UVCS observations on 12-14 March 2008; radial distance $r$, polar angle (P.A., counterclockwise from the north pole), beginning and end of the observation, spectral line, exposure time, and slit width.
}
\label{table1}
\begin{tabular}{clccccc}     
  \hline                   
$r$  & P.A. & Start obs.  & End obs.& Spectr. line & Exp. time & Slit width \\
 ($R_\odot$)    & ($^\circ$)  & (dd/mm--hh:mm)  & (dd/mm--hh:mm) & & (s) & ($\mu$m) \\
  \hline
1.68 & 203 & 12/03--18:42 & 12/03--19:43   & H {\sc i} Ly $\alpha$ & 3600 &100 \\
1.78 & 205  & 12/03--19:44 & 12/03--21:00   & H {\sc i} Ly $\alpha$ & 4500 &100 \\
1.88 & 207  & 12/03--21:02 & 12/03--22:33   & H {\sc i} Ly $\alpha$ & 5400 &100 \\
1.71 & 203  & 12/03--22:35 & 13/03--06:06   & O~{\sc vi} & 26700 &100 \\
1.81 & 205  & 13/03--06:07 & 13/03--14:34   & O~{\sc vi} & 30000 &100 \\
1.91 & 207  & 13/03--14.35 & 13/03--18:38   & O~{\sc vi} & 14400 &100 \\
1.88 & 207  &13/03--18:40 & 13/03--19:56   & H {\sc i} Ly $\alpha$ & 4500 &100 \\
1.98 & 208  &13/03--19:57 & 13/03--22:55   & H {\sc i} Ly $\alpha$ & 10500 &100 \\
1.91 & 207  &13/03--22:56 & 14/03--05:42   & O~{\sc vi} & 23700 &100 \\
2.01 & 208  &14/03--05:43 & 14/03--18:33   & O~{\sc vi} & 45600 &100 \\
  \hline
\end{tabular}
\end{table}

\begin{table}[!h]
\caption{ Details for the UVCS observations on 14-17 March 2008; radial distance $r$, polar angle (P.A.), beginning and end of the observation, spectral line, exposure time, and slit width.
}
\label{table2}
\begin{tabular}{clccccc}     
  \hline                   
$r$  & P.A. & Start obs.  & End obs.& Spectr. line & Exp. time & Slit width \\
 ($R_\odot$) & ($^\circ$)   & (dd/mm--hh:mm)      & (dd/mm--hh:mm)   &  & (s) &($\mu$m) \\
  \hline
1.66 & 315  &14/03--18:59 & 14/03--20:00   & H {\sc i} Ly $\alpha$ & 3600 &100 \\
1.76 & 314  &14/03--20:01 & 14/03--21:17   & H {\sc i} Ly $\alpha$ & 4500 &100 \\
1.86 & 313  &14/03--21:18 & 14/03--22:50   & H {\sc i} Ly $\alpha$ & 5100 &100 \\
1.68 & 314  &14/03--22:51 & 15/03--06:22   & O~{\sc vi} & 26700 &100 \\
1.78 & 313  &15/03--06:24 & 15/03--14:50   & O~{\sc vi} & 30000 &100 \\
1.88 & 312  &15/03--14:52 & 15/03--18:55   & O~{\sc vi} & 14400 &100 \\
1.86 & 313  &15/03--18:56 & 15/03--20:07   & H {\sc i} Ly $\alpha$ & 4200 &100 \\
1.96 & 312  &15/03--20:09 & 15/03--22:41   & H {\sc i} Ly $\alpha$ & 9000 &100 \\
1.88 & 313  &15/03--22:42 & 16/03--05:07   & O~{\sc vi} & 228000 &100 \\
1.99 & 312  &16/03--05:09 & 16/03--17:49   & O~{\sc vi} & 44700 &100 \\
2.06 & 311  &16/03--19:02 & 16/03--20:44   & H {\sc i} Ly $\alpha$ & 6000 &150 \\
2.09 & 311  &17/03--00:52 & 17/03--05:56   & O~{\sc vi} & 18000 &150 \\
  \hline
\end{tabular}
\end{table}

\section{UVCS Observations and Identification of the Structures}

We have analyzed UVCS observations of bright coronal structures performed in the period 12-17 March 2008.
These observations are characterized by a good spatial  coverage  with heliodistances of the slit center from 1.7 to 2.1 $R_\odot$.
The pointing of UVCS was carefully studied, in order to center the structures in the part of the slit which still has the maximum spatial resolution after the O~{\sc vi} detector's electronic problems since January 2006 (L. Gardner, private communication).
We analyse the O~{\sc vi} doublet lines at 1031.93 and
1037.62 \AA\, detected on the UVCS O~{\sc vi} channel (984--1080 \AA), and the H {\sc i} Lyman $\alpha$ line at 1216 \AA\, detected on the same channel following a redundant optical path (Kohl {\it et al.}, 1995).
The slit (the spatial direction of the detector)
is oriented perpendicular to the radial direction defined by the polar angle.
The spectrometer slit was 37 arcmin long, with spatial pixels of 7 arcsec binned in groups of 4 for both the O~{\sc vi} doublet and the H {\sc i} Ly $\alpha$ lines.
 The slit width of the spectrometer,
which determines the spectral
resolution of the observation, was selected to be 100 $\mu$m (corresponding to
28 arcsec and 0.36 \AA) up to 2.0 $R_\odot$~and it was
 150 $\mu$m (corresponding to
42 arcsec and 0,54 \AA) at 2.1 $R_\odot$. 
In order to increase the statistics in the analysis, we have grouped together a number of contiguous exposures at the same height.

\begin{figure}
\centering
\includegraphics[height=5.6cm]{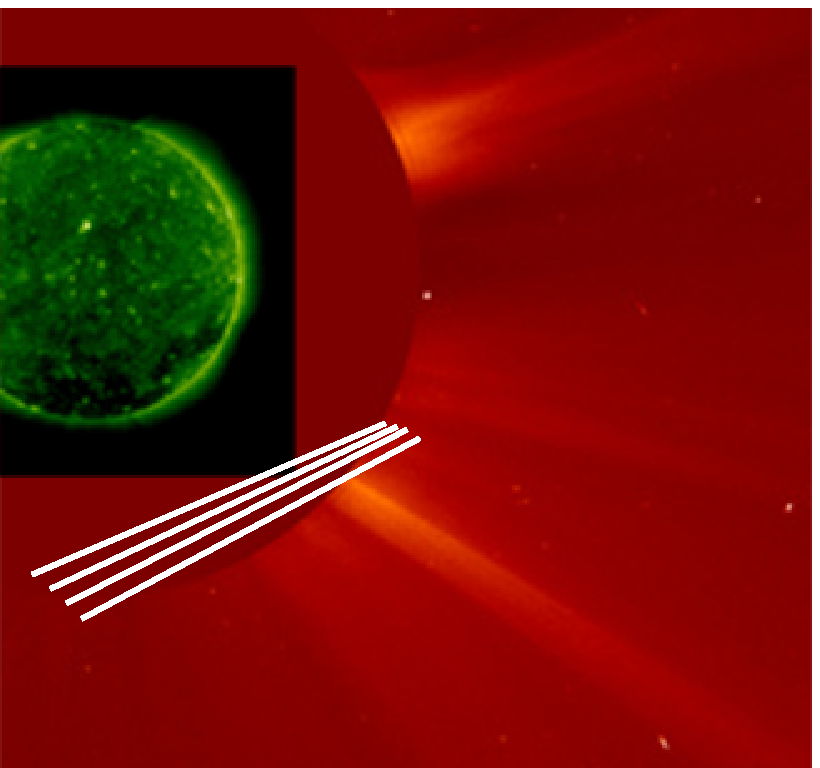}
\includegraphics[height=5.6cm]{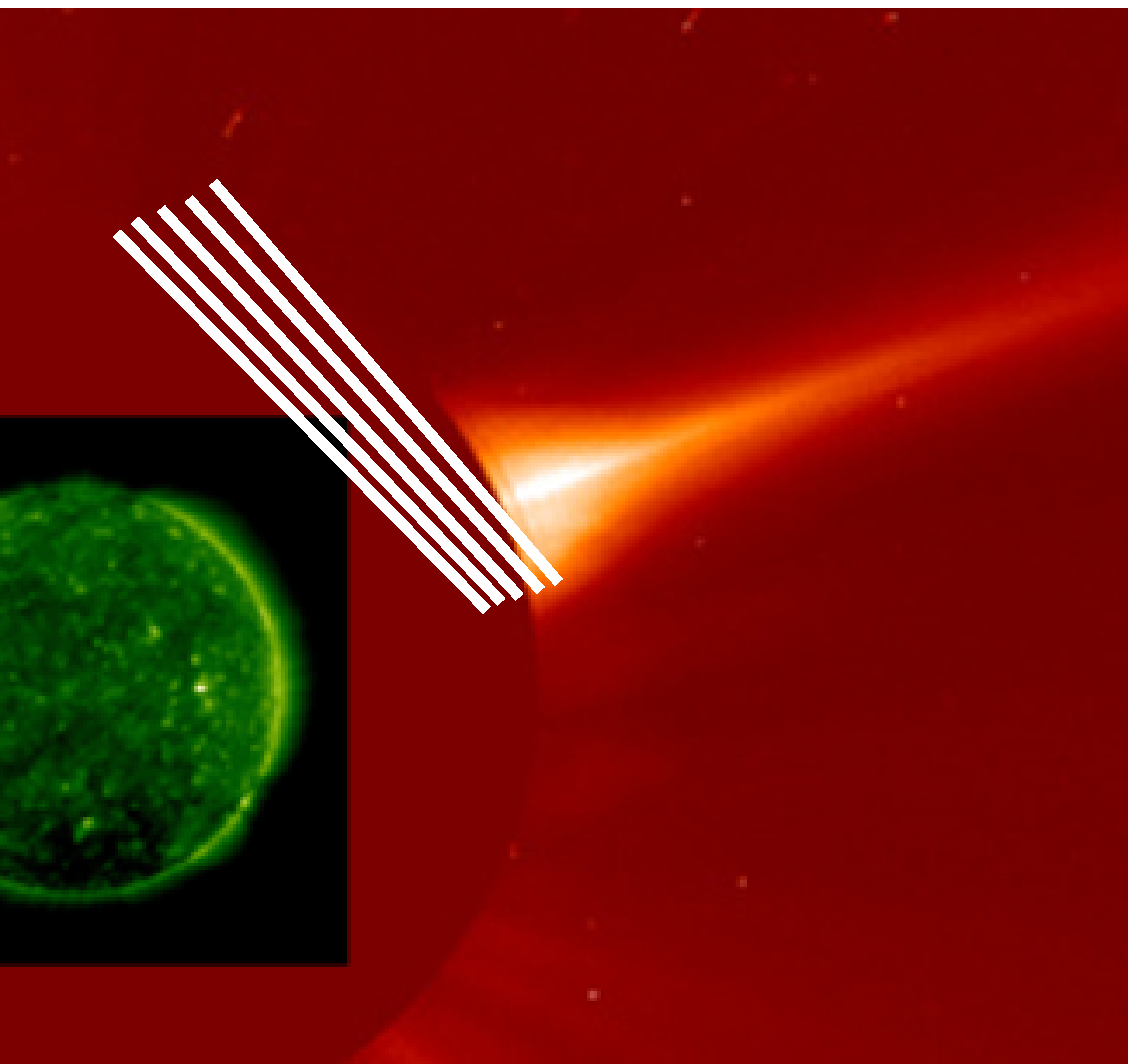}
\caption{Observations on March 2008; disk image from SOHO/EIT 195 {\AA}, visible light image of the corona from SOHO/LASCO C2,  and SOHO/UVCS field of view (white lines) on 12-14 (left) and on 14-17 (right).}
\label{fig1}
\end{figure}

The details of the observations are shown in Table 1 (12-14 March) and Table 2 (14-17 March); radial distance $r$ of the center of the slit, polar angle (P.A., counterclockwise from the north pole), dates of beginning and end of the observation, spectral line, exposure time and slit width.
The {\it Extreme-Ultraviolet Imaging Telescope} (EIT; Delaboudini\`ere {\it et al.}, 1995) 195~\AA~solar disk images are shown in Figure 1 overlapped with the coronal images in visible light from the {\it Large Angle Spectroscopic Coronagraph} (LASCO; Brueckner {\it et al.}, 1995) C2 coronagraph and the field of view of UVCS
 indicated by white lines (left panel for observations on 12-14 March, right panel for observations on 14-17 March).
 In order to identify the kind of streamer observed by UVCS, we have studied the location of the heliospheric current sheet (HCS) which contains magnetic polarity 
reversal and is associated with bipolar streamers, and the plasma sheet (without polarity reversal) associated with pseudo-streamer  structures, by applying a potential-field source-surface (PFSS) extrapolation to photospheric field measurements for Carrington rotation (CR) 2067 (20 February -- 21 March 2008). We have determined the locations and polarities of all open field regions (coronal holes) at the solar surface. 
Figure 2 shows synoptic maps showing the large-scale field and distribution of coronal holes during CR 2067. The top panel shows the photospheric field data from the National Solar Observatory (NSO/Kitt Peak), the middle panel shows the Fe {\sc xv} 284 \AA~emission by the {\it Extreme Ultraviolet Imager} (EUVI; Wuelser {\it et al.}, 2004) on the Ahead (A) spacecraft of the {\it Solar-Terrestrial Relations Observatory} ($STEREO~A$; Kaiser {\it et al.}, 2008), and the bottom panel shows PFSS-derived coronal holes, where the polarity of the underlying photospheric field is indicated by dark gray (if $B_r <$0) or light gray (if $B_r >$0).
The colored dots in the bottom panel represent footpoints of open field lines and are coded according to the associated expansion factors at the source surface (see the figure caption for the numerical values). It is clear that at the longitude and latitude corresponding to the observations at south-west limb (of about $180^\circ$ and $-30^\circ$, respectively) there is an equatorward extension of the south polar hole where the source surface values of the expansion factor are very low.
In this region, a pseudo-streamer is present. This is also supported by Figure 3, where the top panel shows the locations of the HCS/helmet streamer (white pixels) and of the plasma sheet/pseudo-streamers (gray pixels), derived from a PFSS extrapolation of a photospheric field map from the Mount Wilson Observatory. The middle panel shows the white-light observations
above the west limb at 10 $R_\odot$, with the COR2 coronagraph on STEREO-A and the bottom panel illustrates the corresponding simulated white-light patterns. The method for generating the simulated brightness is based on Thomson scattering of photospheric radiation from the helmet-streamer and pseudo-streamer plasma sheets and is described in detail by Wang {\it et al.} (2014). On the one hand, we can see that the white-light structure observed in the northern hemisphere at longitudes $\approx$90-180$^\circ$ is likely the bipolar streamer observed by UVCS at the west limb on 14--17 March; on the other hand, the white-light structure observed near longitude 180$^\circ$ in the southern hemisphere is definitely a pseudo-streamer observed by UVCS at the west limb on 12--14 March.

\begin{figure}
\centering
\includegraphics[height=11cm]{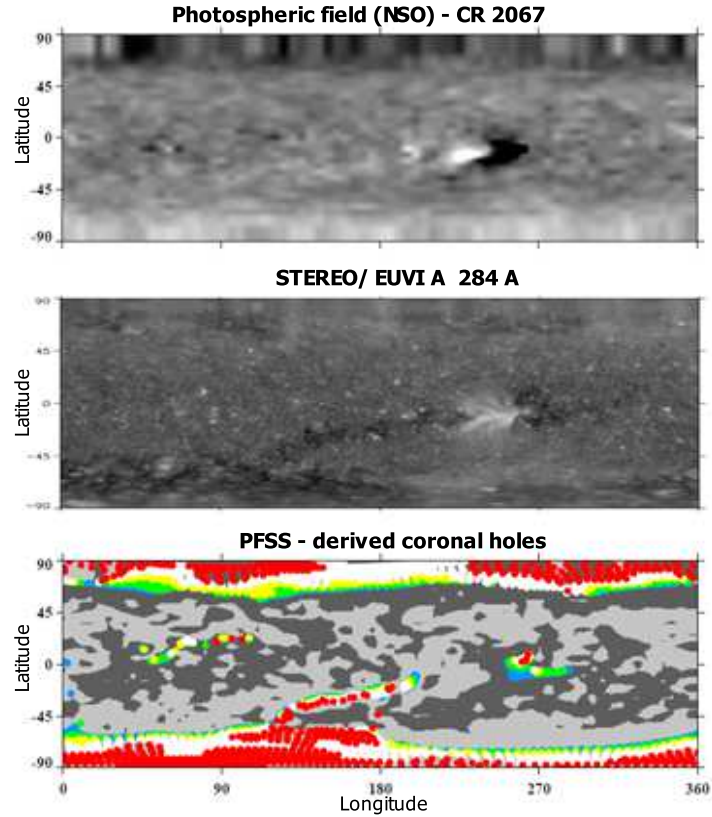}
\caption{Synoptic maps of CR 2067 between 23 February and 21 March 2008 showing NSO/Kitt Peak photospheric magnetic field (top), Fe {\sc xv} 284 \AA~emission recorded by the STEREO-A/EUVI, and PFSS-derived coronal holes (bottom). In the bottom panel, colored dots represent footpoints of open field lines and are coded according to the associated expansion factors at the source surface as $f_{\rm ss} > 20$ (blue), $10 < f_{\rm ss} < 20$ (green), $7 < f_{\rm ss} < 10$ (yellow), $4.5 < f_{\rm ss} < 7$ (white), and $f_{\rm ss} < 4.5$ (red). The polarity of the underlying photospheric field is indicated by dark gray (if $B_r <$0) or light gray (if $B_r >$0).}
\label{fig2}
\end{figure}

\begin{figure}
\centering
\includegraphics[height=11cm]{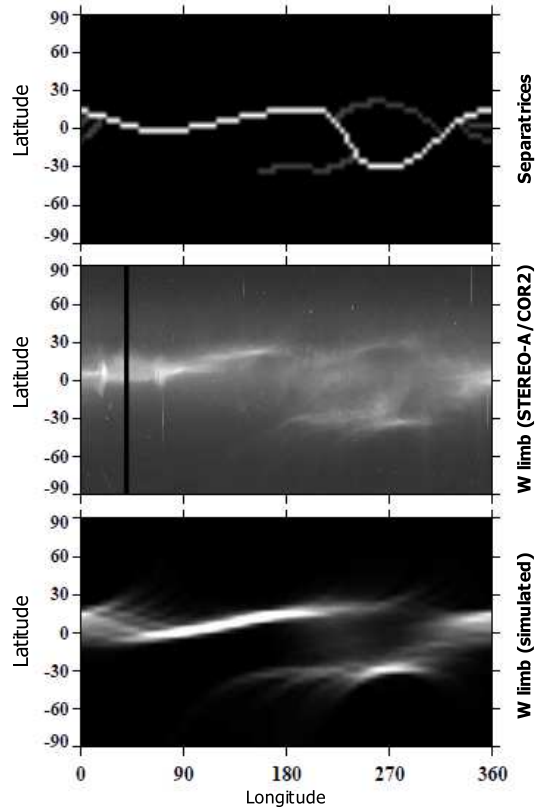}
\caption{Observed and simulated coronal structures during CR 2067. Top panel: Boundaries between open field regions at $r= R_{\rm SS} = 2.5 R_\odot$, derived by applying a PFSS extrapolation to the photospheric map. White pixels represent the source-surface neutral line (or predicted position of the HCS and helmet-streamer plasma sheet), and gray pixels mark the boundaries between coronal holes of the same polarity (or predicted locations of pseudostreamer plasma sheets). Middle Panel: White-light observations above the west limb at 10 $R_\odot$, with the STEREO-A/COR2 coronagraph. Bottom panel: Simulated brightness patterns above the west limb. }
\label{fig3}
\end{figure}

\section{Extrapolations of the Coronal Magnetic Field}

Since we want to determine the structure of the coronal magnetic field for CR 2067 and to 
derive the expansion factors of the flow tubes corresponding to the UVCS observations,
we have used MAS (Magnetohydrodynamics outside A Sphere), the three-dimensional magnetohydrodynamics (MHD) model of Linker {\it et al.} (1999) and Miki\`c {\it et al.} (1999), in
the so-called polytropic approximation of the energy equation
(the model is also capable of calculating the plasma properties solving a more
realistic energy equation; {\it e.g.} Lionello {\it et al.}, 2009).  Starting from a prescribed
magnetic flux distribution at the base of the computational domain, the code integrates the
time-dependent, full MHD equation in spherical coordinates $(r,\theta,\phi)$ until
the configuration reaches a steady state. For the present case, the magnetic field
data has been derived from synoptic observations at National 
Solar Observatory (Kitt Peak) on the days of observation considered in the analysis.
Beside the magnetic flux distribution, to solve the full MHD equations we need to prescribe at the solar surface  the fixed values of temperature and  density, and the components of the velocity.  The latter are determined by solving the gas characteristic equations along magnetic field lines as Miki\`c {\it et al.} (1999). The same equations are also
solved to determine all the properties at the outer boundary, where the flow
is supersonic and super Alfv\'enic. Pseudostreamers are readily identified in MAS,
by tracing the associated magnetic field lines, which do not exhibit a polarity
reversion.
The coronal magnetic field line map extrapolated through the 3D MHD model is shown in Figure 4 where we have labeled pseudo-streamer  as 'PS' at the southern-west limb and a bipolar streamer as 'BS' at the north-west limb.
The pseudo-streamer presents open magnetic field lines while the bipolar streamer is characterized by closed field lines in the region observed by UVCS.

\begin{figure} [h]
\centering
\includegraphics[height=5.6cm]{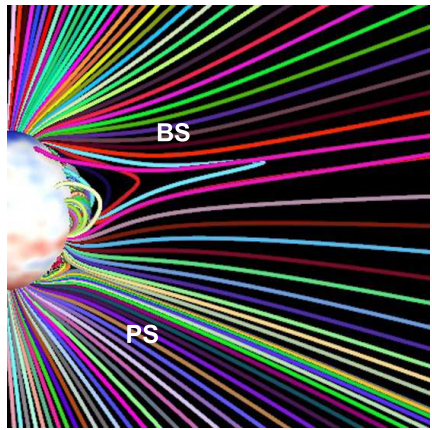}
\caption{Extrapolations of coronal magnetic field on 13 March 2008. We have identified a pseudo-streamer at the south-west limb and a bipolar streamer at the north-west limb indicated in the figure with the labels 'PS' and 'BS', respectively.}
\label{fig4}
\end{figure}

\section{Data Analysis}

 In order to derive the plasma conditions of the coronal structures,
the intensities of the spectral lines are integrated in the two regions defined between $230^\circ$ and $233^\circ$ (counterclockwise from the north pole, 84 arcsec along the slit) for the pseudo-streamer and between $285^\circ$ and $300^\circ$ (counterclockwise from the north pole, 616 arcsec along the slit) for the bipolar streamer. Stray light correction is applied and counts are transformed to intensity, $I(\lambda)$, by applying a radiometric calibration developed by the UVCS group of the Harvard-Smithsonian Center for Astrophysics (J. Kohl, private communication). It has been applied an ad-hoc procedure for these observations and for the Whole Heliosphere Interval, an international campaign to study the three-dimensional solar-heliospheric-planetary connected system near solar minimum which started on 20 March 2008, just after the observations that we planned and here presented analysed.
  The integrated emissions are then fitted with a Gaussian function, representing the coronal profile,
  convolved with a Lorentzian curve which accounts for the instrumental broadening and a rectangular
  function accounting for the width of the spectrometer slit.
  The function resulting from the convolution is added to a background linearly dependent on wavelength.
  The best fit is obtained by applying the least square method, deriving the observed line intensity  as the integral over  the Gaussian line profile.
The electron density and the outflow velocity are derived, as discussed in the previous section, by applying the method which is based on the hypothesis of the expansion factors of the flux tubes connecting the corona and heliosphere. We have obtained these values from the extrapolations of the coronal magnetic fields of the MHD model. In Figure 5, the red solid line corresponds to pseudo-streamer, compared with those derived by the same MHD code for a bipolar streamer boundary in 1996 (cyan dashed line), for the bipolar streamer boundary in 2008 (blue dotted line) and for the northern coronal hole in 2008 (green dash-dotted line). It is evident that the expansion factors of pseudo-streamer are much lower than the values for bipolar streamer, of a factor 2-3. They are closer to coronal hole values but with a different derivative in the height range 1.5-2.5 $R_\odot$, which is the region of UVCS observations.
\begin{figure}[h]
\centering
\includegraphics[height=8cm]{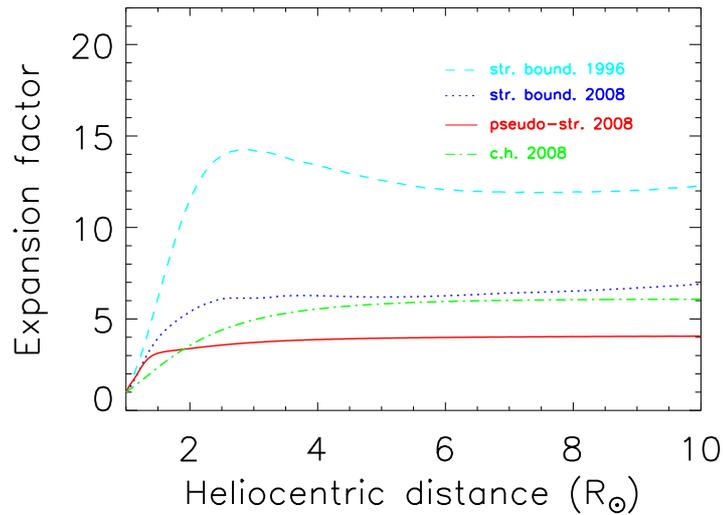}
\caption{Expansion factors as a function of heliocentric distance as derived by the MHD model for the south-western pseudo-streamer of this study (red solid line), for the north-western streamer boundary (blue dotted line), for the northern coronal hole in 2008 (green dash-dotted line) during the same Carrington rotation, and for a streamer boundary observed in 1996 (cyan dashed line).}
\label{fig5}
\end{figure}

Moreover, as already described by Antonucci {\it et al.} (2005) and by Abbo {\it et al.} (2010), the distribution of the
oxygen ion velocity in three dimensions has to be assumed, since, in a given volume, the absorption of the photons
 of the exciting spectrum is controlled by the coronal absorption profile along the incident direction in a solid
 angle subtending the disk of the Sun. The solar wind is assumed to be radial and the velocity
 distribution along the two directions perpendicular to the radial is considered to be the same.
The kinetic temperature of oxygen ions, $T_{\rm k}$, expressed in terms of the spectral line width observed by UVCS, is a measure of the velocity distribution width
along the LOS. Some assumptions are needed for the perpendicular directions to the LOS;
an isotropic Maxwellian velocity distribution with the width defined by the observed $T_{\rm k}$, and a bi-Maxwellian velocity distribution
 of the ions. In this last case, the ion kinetic temperature corresponds to the observed line width in the plane perpendicular
 to the radial direction, and the ion kinetic temperature along the radial direction is equal to the electron temperature (maximum anisotropy).
The isotropic hypothesis is applied inside the bipolar streamer since the electron and ion densities are higher and due to the approximately static conditions
of the plasma, the isotropy of the ion velocity distribution is established via ion-ion collision.
For the pseudo-streamer, we assume both an isotropic and anisotropic velocity distribution since  
they represent an intermediate condition between closed field regions and the core region
of coronal holes, where the ion velocity distributions are found to be highly anisotropic
 ({\it e.g.} Cranmer {\it et~al.}, 1999; Antonucci {\it et~al.}, 2000).
 The coronal electron temperature, $T_{\rm e}$, assumed in the analysis, is that
derived by Gibson {\it et~al.} (1999) for the minimum of solar activity and varies between 1.5$\times 10^6$ K and 9.5$\times 10^5$ K in the range of distance 1.5--2.5 $R_\odot$.
 We point out that
the $T_{\rm e}$ values do not influence significantly the results of the analysis
of electron density and outflow velocity: by assuming the electron temperature values of coronal hole, the obtained results are within the errors shown in Figures 7 and 8 (see the following section).

\section{Results and Conclusions} 
The coronal plasma physical parameters have been derived for a pseudo-streamer and a bipolar streamer which are now compared with those of a bipolar streamer and of a coronal hole observed by SOHO/UVCS in 1996 and already published by Abbo {\it et al.} (2010) and by Antonucci {\it et al.} (2000), respectively.
The kinetic temperature of ions, $T_{\rm k}$, expressed in terms of the spectral line width observed by UVCS,  $\sigma_{\lambda}$,
 is a measure of the velocity distribution of the ions
along the line of sight and can be written as $T_{\rm k} = \frac{A_{\rm i} m_{\rm p}}{k_{\rm B}} \frac{c^2}{\lambda_0^2} \sigma_{\lambda}^2 $, where $A_{\rm i}$ is the ion mass number, m$_{\rm p}$ is the proton mass, $k_{\rm B}$ is the Boltzmann constant, c is the light speed and $\lambda_0$ is the center wavelength of the spectral line. The width of the velocity distribution of the atoms/ions is the result primarily by thermal motions. However, also other motions of the atoms/ions can be induced by plasma waves, turbulence and other microscopic or bulk motions along the line of sight. The kinetic temperature is a quantity measuring all motions of both thermal and non-thermal origin.

Figure 6 shows the kinetic temperature of H {\sc i} atoms (triangles) and O {\sc vi} ions (full dots) as a function of heliodistance for pseudo-streamer (red), streamers (blue for 2008 data-set and cyan for 1996 data-set) and coronal hole (green). The values of neutral hydrogen and O {\sc vi} ions in the coronal hole are from Antonucci {\it et al.} (2000), while the values in streamer on 1996 are from Abbo {\it et al.} (2010).
It is worth noting for the pseudo-streamer results that, on one hand, the $T_{\rm k}$ of H {\sc i} Ly$\alpha$ values  slightly decrease starting at 2.1 $R_\odot$; on the other hand the oxygen ion kinetic temperatures show a rapid increase from 2.1 $R_\odot$. The broadening of the spectral lines can be a signature of energy deposition in the extended corona, which causes the solar wind acceleration, as suggested by the interpretation of coronal hole observations ({\it e.g.} Antonucci {\it et al.}, 2000), and likely it is what happens also in pseudo-streamers. 
 We have also derived the electron density for pseudo-streamer and bipolar streamer and the results are shown in Figure 7 as a function of heliocentric distance. The electron density values relative to the bipolar streamers are shown as dots for the 1996 and 2008 data set, in color cyan and blue respectively. The results for pseudo-streamer are shown as a red region which includes the values obtained with the anisotropic and isotropic hypothesis of the ion velocity distribution (see previous section) and they are intermediate between those derived for streamers by Gibson {\it et~al.} 
(1999; dashed line) and  for coronal holes by Guhathakurta {\it et~al.} 
(1999; dotted line).
The n$_{\rm e}$ values of bipolar streamer are comparable with those derived by  Gibson {\it et~al.} (1999).
 
\begin{figure}
\centering
\includegraphics[height=7cm]{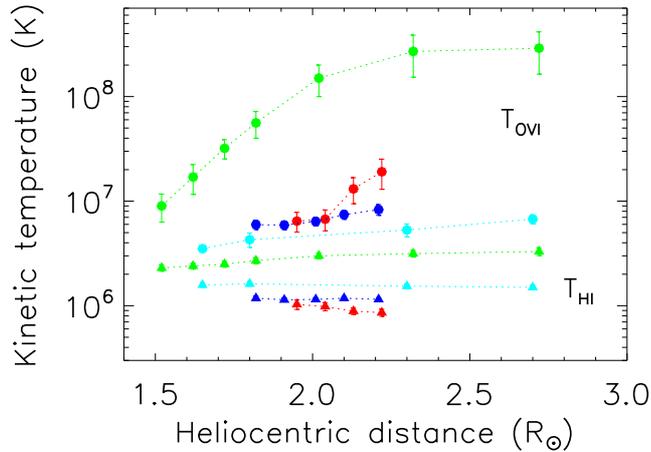}
\caption{Kinetic temperature of H {\sc i} Ly$\alpha$ (triangles) and O {\sc vi} 1032 {\AA} (full dots) as a function of heliocentric distance for pseudo-streamer (red), streamers (blue for 2008 and cyan for 1996), and coronal hole (green). See the text for references.}
\label{fig6}
\end{figure}

\begin{figure}
\centering
\includegraphics[height=7cm]{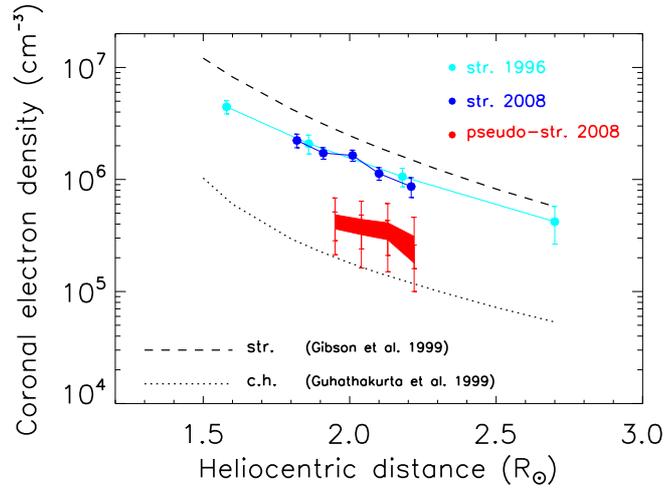}
\caption{Electron density as a function of heliocentric distance relative to pseudo-streamer (red region) and bipolar streamers (blue for 2008 data-set, cyan for 1996 data-set). The dashed line shows the values derived by Gibson {\it et~al.}  (1999) from visible light observations for streamers and the dotted line shows results obtained by Guhathakurta {\it et~al.} (1999) for coronal holes.}
\label{fig7}
\end{figure}

\begin{figure}
\centering
\includegraphics[height=8cm]{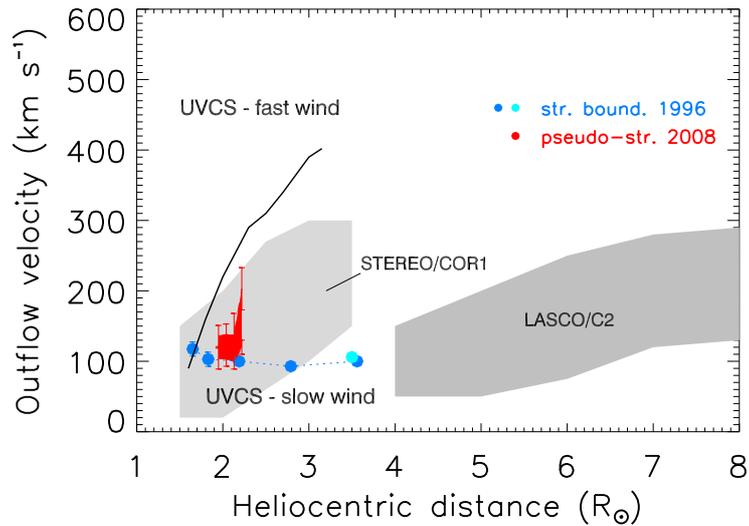}
\caption{Outflow velocity as a function of heliocentric distance relative to pseudo-streamer (red region), compared to the bipolar streamer at the boundary with coronal hole (light blue dots) and at the cusp (cyan dot). As reference, the outflows of a coronal hole are shown as solid line and the grey
bands show the range of outflow velocities for the slow
wind obtained with SOHO/LASCO C2 in 1996 from 4 to 8 \rsun~ and with STEREO/COR1 from 1.5 to 3.5 \rsun~ (see the text for references).}
\label{fig8}
\end{figure}

 The outflow velocity values found in the pseudo-streamer are in the range 105-200 km s$^{-1}$ from 1.95 to 2.22 \rsun~ and they are shown in Figure 6 by the red region which includes the values obtained with the anisotropic and isotropic hypothesis of the ion velocity distribution. These results are compared with those obtained from a bipolar streamer on 1996 at the boundary with coronal hole (light blue dots) and at the cusp (cyan dot). For the bipolar streamer on 2008, we assume a static condition since the analysed structure is located within the closed magnetic field lines region, inside the helmet streamer cusp.  
As reference, the outflows of a coronal hole are shown as solid line (derived by UVCS observations; Antonucci {\it et al.},  2000) and the grey
bands show the range of outflow velocities of 'blobs'/inhomogeneities, tracers for the slow
wind, obtained with LASCO C2 in 1996 from 4 to 8 \rsun~(Sheeley {\it et al.}, 1997) and with STEREO/COR1 from 1.5 to 3.5 \rsun~(Jones and Davila, 2009). 
The results of the analysis show some peculiarities of the pseudo-streamer physical parameters in comparison with those obtained for bipolar streamers. In particular, we have found higher kinetic temperature, higher outflow velocities of O {\sc vi} ions and lower electron density values.
In conclusion, pseudo-streamers are coronal structures much studied in recent years, and the subject of debate as to their contribution to the fast or slow solar wind. 
Empirically based models applied to pseudo-streamers give different results: in particular, according to the boundary layer model of slow wind (interchange reconnection between open and closed field lines at coronal hole boundaries), pseudo-streamers should present a slow flow, while for the expansion factor model (related to the geometrical properties of flux tubes as they expand into the heliosphere) they have a faster flow, even faster than wind originating from coronal holes (Riley {\it et al.}, 2012).
Panasenco and Velli (2013) pointed out that, from a global magnetic configuration reconstructed with PFSS models, the expansion factor of pseudo-streamers  does not increase monotonically with heliocentric distance, but depend on the entire 3D magnetic field configuration; hence, these structures can originate fast or slow wind.
From our study, we point out that pseudo-streamers produce a ''hybrid'' type of outflow that is intermediate between slow and fast solar wind, in according with the in-situ observations of velocity and elemental composition (Wang {\it et al.}, 2012), and they are a possible source of slow/fast wind in not dipolar solar magnetic field configuration.


\begin{acks}
We thank John Kohl and Larry Gardner for helping us in the careful calibration of the UVCS data. UVCS is a joint project of the
National Aeronautics and Space Administration (NASA), the Agenzia
Spaziale Italiana (ASI) and Swiss Founding Agencies.
We thank also SOHO/EIT and LASCO, STEREO/EUVI and COR2, NSO/Kitt Peak, Mt. Wilson Observatories for use of their data.
The research of LA has been funded through the contract
 I/023/09/0 between the National Institute for Astrophysics (INAF)
 and the Italian Space Agency (ASI).   
\end{acks}

\end{article}

\end{document}